# Tensor Completion with Provable Consistency and Fairness Guarantees for Recommender Systems

Tung D. Nguyen and Jeffrey Uhlmann

Dept. of Electrical Engineering and Computer Science
University of Missouri - Columbia

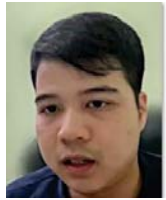
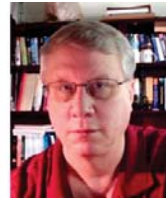

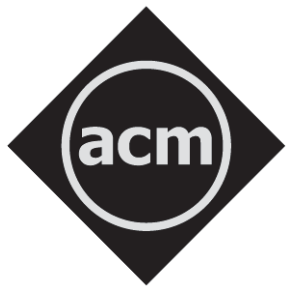

# Tensor Completion with Provable Consistency and Fairness Guarantees for Recommender Systems

TUNG NGUYEN and JEFFREY UHLMANN, University of Missouri, USA

We introduce a new consistency-based approach for defining and solving nonnegative/positive matrix and tensor completion problems. The novelty of the framework is that instead of artificially making the problem well-posed in the form of an application-arbitrary optimization problem, e.g., minimizing a bulk structural measure such as rank or norm, we show that a single property/constraint: preserving unit-scale consistency, guarantees the existence of both a solution and, under relatively weak support assumptions, uniqueness. The framework and solution algorithms also generalize directly to tensors of arbitrary dimensions while maintaining computational complexity that is linear in problem size for fixed dimension $d$. In the context of recommender system (RS) applications, we prove that two *reasonable* properties that should be expected to hold for any solution to the RS problem are sufficient to permit uniqueness guarantees to be established within our framework. This is remarkable because it obviates the need for heuristic-based statistical or AI methods despite what appear to be distinctly human/subjective variables at the heart of the problem. Key theoretical contributions include a general unit-consistent tensor-completion framework with proofs of its properties, e.g., consensus-order and fairness, and algorithms with optimal runtime and space complexities, e.g., $O(1)$ term-completion with preprocessing complexity that is linear in the number of known terms of the matrix/tensor. From a practical perspective, the seamless ability of the framework to generalize to exploit high-dimensional structural relationships among key state variables, e.g., user and product attributes, offers a means for extracting significantly more information than is possible for alternative methods that cannot generalize beyond direct user-product relationships. Finally, we propose our consensus ordering property as an admissibility criterion for any proposed RS method.

**ACM Reference format:**
Tung Nguyen and Jeffrey Uhlmann. 2023. Tensor Completion with Provable Consistency and Fairness Guarantees for Recommender Systems. *ACM Trans. Recomm. Syst.* 1, 3, Article 15 (August 2023), 26 pages.
https://doi.org/10.1145/3604649

T. Nguyen and J. Uhlmann contributed equally to this research.
The first author would like to thank the Electrical Engineering and Computer Science (EECS) Department of the University of Missouri-Columbia for an Undergraduate Research Stipend that helped to support this work.
Authors' address: T. Nguyen and J. Uhlmann, University of Missouri, 201 Naka Hall, Columbia, Mo, USA, 65211; emails: tdn84d@mail.missouri.edu, uhlmannj@missouri.edu.

## 1 INTRODUCTION

Matrix/tensor completion is inherently ill-posed and thus demands additional constraints to ensure the existence of a nonempty and nontrivial family of solutions. Historically the focus has been on *the matrix* (or *the tensor*) as a mathematical object rather than on properties that should be preserved based on the geometry or other structural constraints implicit to the application. Consequently, the conventional means for defining matrix completion is to formulate it as an optimization problem such as minimizing a specified norm of the matrix or tensor [6, 7, 12, 15–17, 19, 30, 35, 37]. At a minimum, however, the choice should be consistent with properties required by the application. For example, if the geometric structure of the application's solution space is understood to be rotation-invariant, then clearly a unitary-invariant measure of error would be appropriate. On the other hand, if the application involves variables with incommensurate units of measure (e.g., length and pressure variables defined in arbitrary metric or imperial units), then minimizing a unitary-invariant norm is meaningless because solutions will be dependent on the arbitrary choice of units made for the application's state variables [45, 46].

We take the position that a well-posed formulation of the matrix/tensor completion problem should derive from properties of the solution space that must be enforced or conserved. In particular, we assume that the vast majority of nontrivial real-world problems involve some number of known and unknown variables with incommensurate units. Therefore the solution for a given problem must be consistent with respect to those units in the sense that arbitrary relative linear scalings of them, e.g., changing from meters to centimeters for lengths, should yield the same unique solution *but in the new units*.[1] We refer to this as a **unit-consistent** (**UC**) solution [38]. To intuitively appreciate the significance of this perspective, consider the alternative in which a change from millimeters to centimeters yields a fundamentally different solution, i.e., is not the same solution given in the new units. Which solution is "better"? If some meta-criterion selects one choice of units over the other, then does that mean another choice of units might yield an even better solution [45]?

In many areas of applied mathematics and engineering, the default reflex when faced with an ill-posed problem (e.g., an underdetermined set of equations) is to apply a toolbox RMSE method to define a unique solution [2, 13, 40]. The inclination to resort to generic RMSE to impose a solution is so strong as to go almost unnoticed, as if there is no need to reflect on whether the resulting solution is meaningful in terms of the given application. An alternative domain-agnostic approach is to try to identify a formulation of the ill-posed problem so as to minimize the computational complexity of the algorithm needed to solve that formulation. Such an approach implicitly presupposes that there is no particular "best" solution from the perspective of the application, so computational efficiency becomes the principal consideration.

For some practical applications, it could be the case that computational considerations must supersede all else because *some* solution is better than *no* solution. This may motivate a rank-minimization formulation of the matrix completion problem as a means for producing a decomposable (factorable) result that can reduce problem complexity for subsequent operations [10, 11, 14, 22, 31, 34, 36, 41, 42]. But will the resulting solutions be unitary-invariant or unit-consistent? The application may demand one of these (or some other property or properties) be enforced, and identifying which is appropriate should precede consideration of possible problem formulations if only to characterize limitations of the ultimately chosen solution method.

---

[1]Examples include scale-invariant or unit-consistent PCA, MDS, neural networks and machine learning methods, etc., to complement or replace conventional least-squares or other norm-specific criteria. See [44–46] for use of this in the context of robot-control applications, and [1, 3] for use in image interpolation and bioinformatics.

The first contribution of this paper is a fully general and computationally efficient method for transforming a given tensor to a scale-invariant canonical form. This form ensures that whatever operation is applied to it will be scale invariant. For example, a matrix function $f(A)$, can be made scale-invariant as $f(\mathcal{S}(A))$, where the function $\mathcal{S}$ ensures $\mathcal{S}(DAE) = \mathcal{S}(A)$ is uniquely determined and holds for all strictly positive diagonal matrices $D$ and $E$. Alternatively, the function $f$ can be made unit-consistent as $f(DAE) = D f(A) E$, where $D$ and $E$ can be interpreted as defining units on state variables. In summary, the unique canonical scaling $\mathcal{S}(A)$ is invariant with respect to positive diagonal scalings of its matrix argument, and this canonical scaling generalizes to arbitrary $d$-dimensional tensors.

The second contribution of this paper is a tensor completion algorithm based on generalized unit-scale invariant canonical form. We argue that human/subjective variables that are presumed to be unknowable but critical to effective **recommender system (RS)** solutions can be effectively understood as a lack of knowledge of units that are implicitly applied when humans rate products, e.g., that humans implicitly rank using subjective units that may depend on the product (or its class or attributes). Evidence for this conclusion comes from the fact that a unit-consistency constraint on its own is sufficient to perform comparably to state-of-the-art specialized RS systems across a range of crude/arbitrary metrics such as RMSE. Potentially more important from an analytic perspective, the UC constraint alone also determines a unique solution (assuming full support, which will be discussed), and from it we are able to prove a *consensus ordering* theorem that any admissible RS solution should be expected to satisfy: if all users agree on a rank-ordering of a set of products, then recommendations (entry completions) will also satisfy that ordering.

The complexity to transform to scale-invariant form is $\Omega(n)$, where $n$ is the number of known entries. However, our most general formulation of the RS problem for tensors can involve constructions for which each user is represented as a vector of attributes (and products are similarly generalized) and each of these attributes can in principle be recursively expanded to capture whatever relationships are deemed to be of predictive relevance. In practice, one can expect $n$ to be relatively manageable because the collection of information is practically bounded even if the implicit index space becomes exorbitant, but the $O(1)$ per-entry completion complexity, which assumes fixed $d$, can be compromised because we permit predictions to be defined across arbitrary $k$-dimensional subtensors, i.e., our generalization can allow for formulations with $O(\binom{d}{k})$ entry-completion complexities [27]. Fortunately, RS applications of interest will typically involve formulations for which $k = d - 1$, thus preserving the optimal complexities achieved in the matrix case to arbitrary $d$-dimensional tensors.

The format of the paper is as follows. We begin by introducing the recommender system problem and arguing that unit consistency can serve as a defining solution property. We then formally define a general **canonical scaling algorithm (CSA)** for tensors, followed by the tensor completion algorithm based on that canonical scaling, which provides a direct completion method for recommender-system applications. We then prove general UC properties of the completion algorithm, e.g., relating to uniqueness, followed by proofs of properties that are of specific relevance to recommender systems. Notably, we prove a consensus-ordering theorem, which essentially says that if all users agree on an ordering-by-preference for a set of products, the estimated ratings (recommendations) obtained from the completion algorithm will respect that ordering. We also generalize the interpretation of the consensus-ordering theorem to apply with respect to user attributes and/or product attributes and/or the tensor extension of any other state variable. Lastly, we show that the approach satisfies a precise notion of *fairness*. We conclude with a discussion of results and their implications for generalized recommender systems.

## 2 RECOMMENDER SYSTEMS

A motivating application for our **unit-consistent (UC)** framework is recommender systems in which a table (matrix) of user ratings of products is used to infer values for unfilled entries, i.e., predict ratings for particular products not rated by particular users. For a matrix $A$ with user $i$ and product $j$, we consider the rating of user $i$ for product $j$ as $A_{ij} = A(i, j)$ for $(i, j) \in \sigma(A)$. We define the list of given/known entries in $\sigma(A)$ as $A_r$, and the list of missing/absent entries in $\overline{\sigma}(A)$ as $A_{nr}$. Thus, the RS recommendation process uses the entries $A_r$ to determine entries for $A_{nr}$. Our goal is to formulate a recommender process such that the resulting entries, $A_{nr}$, satisfy unit consistency and consensus-ordering constraints.

Our approach involves transforming the matrix of user-product ratings to a scale-invariant canonical form by applying a left and right diagonal scaling so that the product of nonzero entries in each row and column is unity. The choice of unity provably guarantees both existence and uniqueness of the solution. The unit-product constraint also uniquely prescribes 1 as the only value that can replace a missing entry, when considered in isolation, and preserve the canonical form, i.e., unit row and column products. When the canonical form is then transformed back via inverses of the left and right diagonal scalings, the resulting values for absent entries will be structurally unit-consistent with respect to the known entries. In other words, if row $i$ of the original matrix is scaled by a positive value $x$, and column $j$ is scaled by a positive value $y$, then the value generated for unfilled entry $i, j$ will be scaled by a factor of $xy$.

To appreciate the need for RS unit consistency, consider a user Alice who rates products in terms of an implicit personal "unit of quality" that derives in unknowable ways from various aspects of her personality and experience [39]. Suppose the RS suggests a rating of $x$ for a new film she has not yet rated. Now consider an alternative scenario in which Alice's personal "unit of quality" is arbitrarily scaled by a factor of 1.05, i.e., all of her ratings become 5% larger. In this scenario we should expect the RS to suggest a rating of $1.05 \times x$ for the new film, i.e., the same value but now consistent with her ratings using the new unit of measure. If it does not, then the RS is not unit consistent.

Continuing with the example, suppose the RS gives Alice a recommended rating of $x$ for the new film based on her original set of ratings, and that value happens to be the same as the rating she has given to a different film. In other words, the RS has implicitly concluded that she will likely give the same rating to the new film as she gave to the other film, i.e., she will have comparable preferences for them. In the alternative scenario in which all of Alice's ratings are increased by 5%, it is natural to expect that the recommended rating for the new film to Alice should be 5% higher. Otherwise, the scaling of Alice's set of ratings would lead to a result in which the system no longer concludes that she will rate the new film comparably to the other film, which implies that the RS does not enforce unit consistency. Unit consistency would, by definition, produce a new rating that is scaled consistently, i.e., the original RS rating of $x$ for the film would become $1.05 \times x$ in the alternative scenario in which all of Alice's ratings are scaled by 1.05.

A different perspective on the need for unit consistency is to consider the influence of Alice's ratings on the recommendations made by the RS to other users. Unit consistency implies that a fixed scaling of all of Alice's ratings, e.g., by a factor of 1.05, will have no effect on RS-suggested ratings for other users. This is because its predicted ratings are invariant with respect to scale factors applied separately to any row or column of the ratings matrix. This invariance also ensures that no individual user is advantaged or penalized in terms of their influence on system ratings due to the magnitude of their personal "unit of quality" they use (implicitly) to produce their ratings. This notion of *fairness* is examined more formally in Section 4.2.

By contrast, an approach that optimizes RS ratings to minimize a measure of squared error – *and hence cannot provide unit consistency* – will implicitly apply more weight to larger-magnitude user ratings than to smaller ones.[2] This means that users who tend to be more reserved in their ratings will have less influence on the system's ratings than users who tend to give higher ratings on average by some factor. Returning to the earlier example, the scaling of Alice's ratings by 5% in a system that minimizes squared error would have the effect of giving her ranked ordering of films slightly more influence on the system's predicted ratings to other users than it had before. This potential source for manipulation of the RS is avoided under the unit-consistency constraint.

There are, of course, opportunities for users to alter the performance of any RS system because its suggested ratings are necessarily derived from user-provided ratings, so if spurious information is entered into the system, then the validity of system outputs will be compromised. This kind of corruption of results is unavoidable. The question is whether, and the extent to which, users can intentionally manipulate the system in a way that unfairly increases the influence of their preferences on the recommendations the system gives to other users. In the case of a unit-consistent RS, if a user feels that film A is 5% better than film B, and film B is 3% better than film C, then the influence of this degree of relative preferences of the user on system ratings cannot be increased or decreased simply by scaling the ratings, e.g., by scaling the three film ratings by a value such that film A has the highest possible rating. The user can only affect the system's ratings for other users by changing his *relative* preferences for the three films, e.g., by giving them all the highest possible ratings. Doing so will not change the *magnitude* of his influence on system ratings, it will only cause the system to believe he likes all three films equally. In other words, he is constrained in his ability to artificially bias a unit-consistent RS toward ratings that are more aligned with his personal relative preferences for the three films: *the user can only affect what the system assumes to be his relative preferences*. In this sense, unit consistency provides a natural and intuitive form of both robustness and fairness.

In the following sections, we formally describe our unit-consistent framework and its properties relating to recommender systems.

## 3 UC COMPLETION FRAMEWORK

### 3.1 Definitions

At a high level, our RS framework transforms a given set of entries in a matrix or tensor to a unique canonical form for which there exists a unique value that preserves the canonical form for each absent entry. Each absent entry in the canonical form is replaced with that value, and the entire set is transformed back to the original space. At the conclusion of this process, the original absent entries will have values (typically distinct) that respect unit consistency. For example, consider a rating matrix with an absent entry in row $i$ that is replaced with a value $x$. If the process were to be repeated for that entry, but now with row $i$ scaled by a factor of 2, then the replaced value for the absent entry would be $2x$ as required for unit consistency. A key component of the framework, therefore, is the required scaling transformation to a unique canonical form. In this section, we present a **Tensor Completion Algorithm (TCA)** for our general **canonical scaling algorithm (CSA)**, and we formally establish both its correctness and the uniqueness of its result for the **canonical scaling problem (CSP)** involving a $d$-dimensional tensor $A$, where the matrix case with $d = 2$ corresponds to the conventional RS tabular formulation.

We begin by defining our notation:

(1) $A \in \mathbb{R}_{>0}^{n_1 \times \cdots \times n_d}$ is a positive $d$-dimensional tensor with fixed dimensional extents $n_1, \ldots, n_d$. (The restriction to strictly positive, as opposed to nonnegative, entries is unnecessary for

---

[2]Many approaches for recommender systems explicitly minimize squared error through the use of the singular value decomposition (SVD) or implicitly via the use of the Moore-Penrose pseudoinverse [4, 18, 21, 23, 28–30, 43].

canonical scaling but will prove convenient for maintaining consistency with the common RS convention of reserving the value of 0 to represent absent entries, and also for simplifying notation by avoiding the need to define a special symbol to designate absent entries.)

(2) $\vec{\alpha} = \{\alpha_1, \ldots, \alpha_d\} \in \mathbb{Z}^d$ is a $d$-dimensional vector that specifies an entry of $A$ as $A(\vec{\alpha})$.
(3) $\sigma(A) = \{\vec{\alpha} \mid A(\vec{\alpha})\}$ is the set of known/defined entries of tensor $A$; and its complement, $\overline{\sigma}(A)$, is the set of absent/missing entries of $A$.
(4) Integer $k < d$ denotes the dimensionality of a given subtensor.
(5) $[m] = \{1, 2, \ldots, m\}$ is an index set of the first $m$ natural numbers.

The following is the first of two key definitions.

*Definition 3.1 (Sets of Subtensors).* Let $V_i = \mathbb{R}^{n_i}$ and assume a $d$-dimensional tensor $A \in V_1 \otimes V_2 \otimes \cdots \otimes V_d$ and a positive integer $1 \leq k < d$. For a subset $\pi = \{\pi_1, \ldots, \pi_k\} \subset [d]$ of cardinality $k$, a flattening $(\pi, \pi^c)$ of $A$ is the matrix $A_{(\pi^c, \pi)}$ defined as:

$$A_{(\pi^c, \pi)} \in \left(V_{\pi_1^c} \otimes \cdots \otimes V_{\pi_{d-k}^c}\right) \otimes \left(V_{\pi_1} \otimes \cdots \otimes V_{\pi_k}\right)^* \tag{1}$$

where $V_{\pi_i}^*$ denotes the dual space of $V_{\pi_i}$. Observe that the rows of $A_{(\pi^c, \pi)}$ matrix contain all the k-dimensional subtensors corresponding to the subset $\pi$. The set of all subtensors is denoted as $\mathcal{A}$ (with finite cardinality $|\mathcal{A}|$) is:

$$\mathcal{A} = \bigcup_{\pi \subset [d]} \{A_{(\pi^c, \pi)}[r, :] \mid 1 \leq r \leq size(A_{(\pi^c, \pi)}, 1)\} \tag{2}$$

where $size(T, n)$ is the dimensional length of a matrix $T$ at dimension $n$.

The flattening operation representing a tensor product of vector spaces puts the indices corresponding to the dimensions specified by $\pi$ first, and the indices corresponding to the dimensions specified by $\pi^c$ last. Thus, the rows of $A_{(\pi^c, \pi)}$ correspond to the k-dimensional subtensors of $A$, where the first $k$ indices correspond to the dimensions specified by $\pi$, and the last $(d-k)$ indices correspond to the dimensions specified by $\pi^c$. (Modern programming libraries support the flattening transformation of a tensor $A$ using *transpose* and *reshape* operations.)

The following is our second key definition.

*Definition 3.2 (Scaling k-dimensional Subtensors of a Tensor).* For $d$-dimensional tensor $A$, let $S_\pi$ be the vector that scales all $k$-dimensional subtensors by subset $\pi \subset [d]$ with cardinality $k$ and let $S_k := (S_\pi)_{\pi \subset [d]}$. We define the $\pi$-product as

$$S_\pi *_\pi A = diag(S_\pi) \cdot A_{(\pi^c, \pi)} \quad \forall \pi \subset [d]. \tag{3}$$

Then the product $A'' = S_k *_k A$ is defined as

$$(S_{\pi_1}, \ldots, S_{\pi_k}) *_k A := S_{\pi_1} *_{\pi_1} (\cdots (S_{\pi_{k-1}} *_{\pi_{k-1}} (S_{\pi_l} *_{\pi_l} A)) \cdots) \tag{4}$$

By formula (3), any $A_i \in \mathcal{A}$ corresponding to a subset $\pi$ is scaled by a coefficient from vector $S_\pi$. This definition is also equivalent to multilinear multiplication in the case $k = d - 1$.

### 3.2 Special Case $k = d$-1 for Scaling Subtensors

In the case of $k = d$-1, the $d$-1 product is equivalent to multilinear multiplication. Specifically as $\pi \in \{\{1\}, \ldots, \{d\}\}$, the scaling of $k$-dimensional subtensors is

$$(S_{\{1\}}, \ldots, S_{\{d\}}) *_d A = (S_{\{1\}}, \ldots, S_{\{d\}}) \cdot A = (S_{\{1\}} \otimes \cdots \otimes S_{\{d\}}) (A) \tag{5}$$

In this case, we have the following observation

OBSERVATION 3.1. *For a d-dimensional tensor A and $\pi \in \{\{1\},\ldots,\{d\}\}$, the equivalency with Definition 3.2 for each $\vec{\alpha} \in \sigma(A)$ and $k = d$-1 is*

$$A''(\vec{\alpha}) \equiv A(\vec{\alpha}) \cdot \prod_{i=1}^{d} S_{\{i\}}(\alpha_i). \qquad (6)$$

where $S_{\{i\}}(\alpha_i)$ is the $\alpha_i$-th entry of the vector $S_{\{i\}}$.

Furthermore, each subtensor element of $\mathcal{A}$ can be expressed by each dimension $n_i$ from $V_i$ and index $j$ for $j \in [n_i]$. For the matrix case $d = 2$, we have two dimension $n_1$ and $n_2$ representing the dimension of rows and columns and $j$ represent an index or row or column. When $d = 3$, we have three dimensional numbers $n_1, n_2$, and $n_3$ and each corresponding $j$ representing a tensor slice in a fixed dimension $i$. For that, we have the following definition

*Definition 3.3.* For $k = d$-1, we denote a subtensor element from set $\mathcal{A}$ for $A \in V_1 \otimes \cdots \otimes V_d$ by each dimension $n_i$ from $V_i$ and index $j \in [n_i]$ as $A_j^{(i)}$.

Due to their simpler forms, we will primarily use formula (6) and Definition 3.3 for our subsequent analyses. The structured scaling of Definition 3.2, Observation 3.1, and Definition 3.3 are fundamental components of the solution described in the next section for the *canonical scaling problem*, which requires the transformation of a given nonnegative tensor to a unique scale-invariant canonical form.

### 3.3 Tensor Completion Algorithm

Here we provide the tensor completion algorithm and its basis as the CSA algorithm, formulated by the CSP problem.

---
**ALGORITHM 1:** Tensor completion algorithm (TCA)

**Input**: $d$-dimensional tensor A and $d$.
**Output**: $A'$
**Function** TCA(A):
- **Step 1: CSA process.**
  $S_{d-1} \leftarrow \text{CSA}(A)$
- **Step 2: Tensor completion process**:
  for $\vec{\alpha} \in \overline{\sigma}(A)$ do
   $A'(\vec{\alpha}) \leftarrow \prod_{i=1}^{d} S_{\{i\}}^{-1}(\alpha_i).$
  return $A'$.

---

We now present Algorithm 1 for obtaining a unique scaled tensor $A'' = \text{CSA}(A)$.

The time complexity of this algorithm is $O(|\sigma(A)|)$ per iteration, so assuming that $d = O(1)$, which is all but necessary in any realistic practical application, and ignoring polylog factors from the convergence parameter $\epsilon$, then this also represents the overall time complexity of the algorithm. This is because the final step of converting back from the log-space solution is $A(\vec{\alpha}) = \exp(a(\vec{\alpha}))$, which is also $O(|\sigma(A)|)$. A more detailed examination of this time complexity can be found in [27].

In the case of $d = 2$ when $A$ is a matrix, i.e., CSA(A), the set of $k = 1$ subtensors is simply the set of rows and columns. This is a specialized instance of a matrix problem studied in [32] with rows and columns explicitly distinguished for the problem of scaling line products of a matrix to chosen positive values (for which a solution is not guaranteed to exist except in the case we apply with

---

**ALGORITHM 2:** CSA for tensor

**Input**: $d$-dimensional tensor $A$.
**Output**: $A''$ and scaling vector $S_{d-1}$.
**Function** CSA($A$):
- **Step 1: Iterative step over constraints**: Initialize $count \leftarrow 0$, variance variable $v \leftarrow 0$, and let $p$ be a zero vector of conformant length. Let $a$ be the logarithm conversion of $A$, i.e., all known entries are replaced with their logs. Let the log conversion of $S_{d-1}$ as $s_{d-1} = (s_{\{i\}})_{i \in [d]}$.

   **for** *each subtensor $A_j^{(i)}$ with index $i \in [d]$ and $j \in [n_i]$* **do**

$$\rho = -\left[|\sigma(A_j^{(i)})|\right]^{-1} \sum_{\vec{\alpha} \in \sigma(A_j^{(i)})} a(\vec{\alpha})$$

$$a(\vec{\alpha}) \leftarrow a(\vec{\alpha}) + \rho, \quad v \leftarrow v + \rho^2, \quad \text{for} \quad \vec{\alpha} \in \sigma(A_j^{(i)})$$

$$s_{\{i\}}(j) \leftarrow s_{\{i\}}(j) - \rho$$

(7)

- **Step 2: Convergence**: If $v$ is less than a selected threshold $\epsilon$, then exit loop. Otherwise, set $count \leftarrow count + 1$ and return to step 1.

   **return** $A'' = exp(a)$ and $S_{d-1} = exp(s_{d-1})$.

---

line products equal to 1), but from our generalized tensor formulation it can be seen that such a distinction is unnecessary.

We now summarize the theory behind the CSA algorithm, which is the CSP formulation for arbitrary scaling of $d-1$-dimensional subtensors of a given nonnegative tensor $A$, which may be obtained by replacing the elements of a given tensor $A'$ with their magnitudes.[3] Using Definition 3.2 and Definition 3.3, we have:

**Canonical Scaling Problem (CSP):** Find the $d$-dimensional tensor $A'' \in \mathbb{R}_{>0}^{n_1 \times \cdots \times n_d}$ and a positive vector $S_{d-1} = (S_{\{1\}}, \ldots, S_{\{d\}})$ such that $A'' = S_{d-1} *_{d-1} A$ and the product of the known entries of each $(d-1)$-dimensional subtensor $A_j''^{(i)}$ is 1.

Here we state the uniqueness solution of the CSP problem. The proof and additional details can be found in Appendix C.

THEOREM 3.4 (UNIQUENESS OF A"). *There exists at most one tensor $A''$ for which there exists a strictly positive vector $S_{d-1}$ such that the solution $(A'', S_{d-1})$ satisfies CSP. Furthermore, if a positive vector $T_{d-1} = (T_{\{1\}}, \ldots, T_{\{d\}})$ satisfies*

$$\prod_{i=1}^{d} T_{\{i\}}(\alpha_i) = 1 \quad \forall \vec{\alpha} \in \sigma(A), \tag{8}$$

*then $(A'', S_{d-1} \circledast T_{d-1})$ is also a solution, where $\circledast$ is the element-wise product.*

### 3.4 Uniqueness and Full Support

Although the canonical scaled tensor is unique, the scaling vectors, or the resulting $A'$ from the *TCA* process, may not be unless there are sufficient known entries to provide *full support*, which is now defined.

---

[3]The decomposition of a given tensor into the Hadamard product of a nonnegative scaling tensor and a tensor with unit-magnitude negative or complex (or quaternion, octonion, or other basis forms) entries can be thought of as a scale/unit-consistent analog of a polar decomposition. In direct analogy to the positive-definite component of the matrix polar decomposition, the nonnegative scaling tensor developed in this section is unique.

*Definition 3.5.* Given A, we say that tensor is *fully supported* if for every entry $\vec{\alpha} \in \overline{\sigma}(A)$, there exists a nonzero vector $\vec{s}$ such that $(\alpha_1 + \delta_1 s_1, \ldots, \alpha_d + \delta_d s_d) \in \sigma(A)$ for all $2^{d-1}$ choices of $\delta_i \in \{0, 1\}$ such that $\delta_1 + \cdots \delta_d > 0$. We denote $\vec{\alpha}' = (\alpha_1 + \delta_1 s_1, \ldots, \alpha_d + \delta_d s_d) = \vec{\alpha} + \vec{\delta} \cdot \vec{s}$ and the set of such vectors $\vec{\alpha}'$ as $H(\vec{\alpha}, \vec{s})$.

Definition 3.5 essentially says that every unknown entry forms a vertex of a $d$-dimensional hypercube with known entries at the remaining vertices. In the $d = 2$ matrix case, for example, an unknown entry $(i, j)$ must have known entries at $(i+p, j)$, $(i, j+q)$, and $(i+p, j+q)$ for some $p \neq 0$ and $q \neq 0$. In this case, $\vec{\alpha} = (i, j)$ and $\vec{s} = (p, q)$. Thus, full support can be expected to hold even with very high sparsity. From this point forward we will assume full support unless otherwise stated. Using this definition, we obtain the following theorem regarding uniqueness of the recommendation/entry-completion result.

THEOREM 3.6 (UNIQUENESS). *The result from TCA(A) is uniquely determined even if there are distinct sets of scaling vectors $\{S_{d-1}\}$ that yield the same, unique, tensor $A''$ from CSA(A).*

This theorem is the direct corollary from the theorem in the Appendix D in the case $k = d$-1.

### 3.5 Unit Consistency

$A'' = \text{CSA}(A)$ guarantees scale-invariance with respect to every $k$-dimensional subtensor of $A'$. It then remains to show that the completion result from TCA($A$) is unit-consistent.

THEOREM 3.7 (UNIT-CONSISTENCY). *Given a tensor A and an arbitrary conformant positive scaling vector $T_{d-1}$, $T_{d-1} *_{d-1} TCA(A) = TCA(T_{d-1} *_{d-1} A)$, where $T_{d-1}$ scales all $d-1$-dimensional subtensors of A (with operator $*_{d-1}$ as defined in Definition 3.2).*

The theorem is the direct corollary from the theorem in the Appendix E in the case $k = d$-1.

### 3.6 Consensus Ordering Property

We now proceed to prove that our methods provide solutions that satisfy the *consensus ordering property*, which we have earlier proposed to be an admissibility criterion for any RS system. Intuition for this criterion can be established with a simple example. Assume that we have hundreds of users, each of whom has rated hundreds of films. Suppose there exist films X, Y, and Z, and every user has rated X higher than Y, and Y higher than Z. If the recommender system satisfies the consensus ordering property, its recommendations (predicted ratings) to a new user will preserve that rank ordering. In other words, it will not recommend Y over X, or Z over X or Y. A system that produces a predicted rating for Z that is higher than X, when every user in the system has explicitly rated X higher than Z, does not satisfy the consensus ordering property. We consider such a system to be inadmissible.[4]

For any recommender system, we can show that the following method satisfies the consensus ordering property as the sanity check with respect to an ordering relationship of the $(d-1)$-dimensional subtensors of a given $d$-dimensional tensor $A$.

*Definition 3.8 (Set of Consensus Ordering).* For tensor $d$-dimensional tensor $A \in V_1 \otimes V_2 \otimes \cdots \otimes V_d$ and subtensors as defined in Definition 3.3 within dimension $n_d$ and a subset $\gamma \subset [n_d]$ with size $D$, we define a set of known indices $\sigma(\gamma)$ that preserves/follows ordering $\gamma$ in tensor $A$ iff

$$\sigma(A^{(d)}_{\gamma_i}) = \sigma(\gamma), \quad \text{for all } 1 \le i \le D.$$
$$A^{(d)}_{\gamma_a}(\vec{\alpha}) < A^{(d)}_{\gamma_b}(\vec{\alpha}) \quad \forall \vec{\alpha} \in \sigma(\gamma) \text{ and } 1 \le a < b \le D. \tag{9}$$

---

[4]Of course, this admissibility criterion does not apply to rank deviations due to numerical imprecision for a system that provably satisfies it in theory. If a system cannot guarantee consensus ordering in theory, but can provably guarantee it to within a tolerance $\epsilon$, i.e., it can be proved that every predicted rating can be perturbed by an amount less than $\epsilon$ to satisfy consensus ordering, then we propose that the system be referred to as $\epsilon$-*admissible*.

Then the set of absent/missing vectors $\overline{\sigma}(\gamma)$ satisfies $\overline{\sigma}(\gamma) = \overline{\sigma}(A^{(d)}_{\gamma_i})$ for all $1 \le i \le D$.

Now we can formally state the following theorem.

THEOREM 3.9 (CONSENSUS ORDERING). *Given a tensor $A$ and obtained result $A' = TCA(A)$, and subset $\gamma$, $\sigma(\gamma) \ne \emptyset$, then any completion vector $\vec{\alpha} \in \overline{\sigma}(\gamma)$ must satisfy $A'^{(d)}_{\gamma_a}(\vec{\alpha}) < A'^{(d)}_{\gamma_b}(\vec{\alpha})$ when $a < b$.*

PROOF. For any vector $\vec{\alpha} \in \sigma(\gamma)$, the ordering condition from Definition 3.8 gives:

$$A^{(d)}_{\gamma_1}(\vec{\alpha}) < \cdots < A^{(d)}_{\gamma_i}(\vec{\alpha}) < \cdots < A^{(d)}_{\gamma_D}(\vec{\alpha}) \tag{10}$$

Given that $A'$ from TCA is normalized via CSA, then without loss of generality we can omit the dimension $k$ when using notation (4) from Section 2, since $k$ is specified as $k = d - 1$. Then the formulaic relationship between $A$ and $A'$ is analogous as follows:

$$A' = (S_{\{1\}}, \ldots, S_{\{d\}}) *_d A = (I_{\{1\}} \otimes \cdots \otimes I_{\{d-1\}} \otimes S_{\{d\}})(S_{\{1\}} \otimes \cdots \otimes S_{\{d-1\}} \otimes I_{\{d\}})(A) \tag{11}$$

where $I_{\{i\}}$ is the identity matrix. Assuming that vector $S_{d-1} = (S_{\{1\}}, \ldots, S_{\{d\}})$, we obtain the following after the TCA process

$$A^{(d)}_{\gamma_i}(\vec{\alpha}) = S^{-1}_{\{d\}}(\gamma_i) \cdot \prod_{j=1}^{d-1} S^{-1}_{\{j\}}(\alpha_j) \cdot A'^{(d)}_{\gamma_i}(\vec{\alpha}). \tag{12}$$

Substituting into the above inequality, we have:

$$A'^{(d)}_{\gamma_1}(\vec{\alpha}) \cdot S^{-1}_{\{d\}}(\gamma_1) < \cdots < A'^{(d)}_{\gamma_i}(\vec{\alpha}) \cdot S^{-1}_{\{d\}}(\gamma_i) < \cdots < A'^{(d)}_{\gamma_D}(\vec{\alpha}) \cdot S^{-1}_{\{d\}}(\gamma_D) \tag{13}$$

Using the CSA result with respect to $(d-1)$-dimensional subtensor in direction $\gamma_i$:

$$\prod_{\vec{\alpha} \in \sigma(\gamma)} A'^{(d)}_{\gamma_i}(\vec{\alpha}) = 1. \tag{14}$$

Substituting (14) into (13) using the fact that $\sigma(A^{(\gamma_i)}) = \sigma(\gamma)$ for all $1 \le i \le D$ gives

$$S^{-1}_{\{d\}}(\gamma_1) < \cdots < S^{-1}_{\{d\}}(\gamma_i) < \cdots < S^{-1}_{\{d\}}(\gamma_D). \tag{15}$$

For any vector $\vec{\alpha}' \in \overline{\sigma}(\gamma)$, the following entry is uniquely determined from Theorem 3.6,

$$A'^{(d)}_{\gamma_i}(\vec{\alpha}') = S^{-1}_{\{d\}}(\gamma_i) \cdot \prod_{j=1}^{d-1} S^{-1}_{\{j\}}(\alpha'_j), \tag{16}$$

and we therefore deduce that

$$A'^{(d)}_{\gamma_1}(\vec{\alpha}') < \cdots < A'^{(d)}_{\gamma_i}(\vec{\alpha}') < \cdots < A'^{(d)}_{\gamma_D}(\vec{\alpha}'), \tag{17}$$

thus completing the proof. □

## 4 UC APPLICATION FOR RECOMMENDER SYSTEM

In this section, we examine the application of UC completion to recommender systems. Because of its particular relevance to matrix-formulated recommender-system problems, we briefly discuss the special case of $d = 2, k = 1$, for a given $m \times n$ matrix[5] $A \in \mathbb{R}_{>0}^{m \times n}$ with full support. For notational convenience, we define the matrix completion function MCA($A$) as a special case of TCA:

$$\text{MCA}(A) \equiv \text{TCA}(A) \text{ for } d = 2. \tag{18}$$

In this case, the unit-consistency property can be expressed as $R \cdot \text{MCA}(A) \cdot C = \text{MCA}(R \cdot A \cdot C)$ for positive vectors $R$ and $C$, which in terms of matrix multiplication with positive diagonal matrices $R$ and $C$ implies $R \cdot \text{MCA}(A) \cdot C = \text{MCA}(R \cdot A \cdot C)$. The time and space complexity for MCA($A$) is $O(|\sigma(A)|)$.

Theorem 3.9 and Definition 3.8 provide a means to specify the recommendation method in the following definition.

*Definition 4.1.* Denote $RS(\vec{\alpha}) = A'(\vec{\alpha})$ as the recommendation result for vector position $\vec{\alpha}$ having $d$ elements from tensor A with $A' = \text{TCA}(A)$.

### 4.1 Consensus Ordering

Following Definition 3.8 and Theorem 3.9, we formally state the consensus ordering property in the context of a matrix and 3-dimensional tensor.

COROLLARY 4.2 (CONSENSUS ORDERING FOR A MATRIX). *Given a matrix A, MCA(A), set of products $P \subseteq [n]$, and set of users $U \subseteq [m]$. We have the following statements:*

(1) *Given a subset of products $\gamma_P$ and nonempty set $\sigma(\gamma_P)$. Then the recommendation result $RS(u', p)$ for any user-product $(u', p) \in \overline{\sigma}(\gamma_P)$ is unique and follows consensus ordering $\gamma_P$.*
(2) *Given a subset of users $\gamma_U$ and nonempty set $\sigma(\gamma_U)$. Then the recommendation result $RS(u, p')$ for any user-product $(u, p') \in \overline{\sigma}(\gamma_U)$ is unique and follows consensus ordering $\gamma_U$.*

For 3D tensor, we have the following interpretation.

COROLLARY 4.3 (CONSENSUS ORDERING FOR 3-DIMENSIONAL TENSOR). *Given a 3-dimensional tensor $A \in \mathbb{R}_{>0}^{n_1 \times n_2 \times n_3}$. For the recommendation result TCA(A), we have the following statements:*

(1) *Given a subset of users $\gamma_U$ and nonempty set $\sigma(\gamma_U)$. Then the recommendation result $RS(u, \alpha', p')$ for attribute-product vector $(\alpha', p') \in \overline{\sigma}(\gamma_U)$ and each user $u \in U$ is unique and also follows consensus ordering $\gamma_U$.*
(2) *Given a subset of products $\gamma_P$ and nonempty set $\sigma(\gamma_P)$. Then the recommendation result $RS(u', \alpha', p)$ for user-attribute vector $(u', \alpha') \in \overline{\sigma}(\gamma_P)$ and each product $p \in P$ is unique and follows consensus ordering $\gamma_P$.*
(3) *Given a subset of attributes $\gamma_\Gamma$ and nonempty set $\sigma(\gamma_P)$. Then the recommendation result $RS(u', \alpha, p')$ for user-product vector $(u', p') \in \overline{\sigma}(\gamma_\Gamma)$ and each attribute $\alpha \in \Gamma$ is unique and follows consensus ordering $\gamma_\Gamma$.*

The strongest implication of this corollary is the symmetry property of Theorem 3.9 with respect to any dimensions of preference in the tensor: users, products, and attributes, i.e., whether attributes are defined with respect to users or products. The generalization from a 2-dimensional matrix to a 3-dimensional tensor even further exploits this symmetry of ordering with respect to the choice of spatial label on different coordinates.

---

[5]Because all results in this paper are transposition consistent, we implicitly assume without loss of generality that $n \geq m$ purely to be consistent with our general use of $n$ as the variable that functionally determines the time and space complexity of our algorithms.

This corollary sums up how the consensus ordering property with respect to the same index label in the original tensor is preserved in the output tensor. Specifically, the ordering follows trivially from Theorem 3.9 whenever there exists unanimity among attributes on the first $d$-1 dimensions with respect to the last coordinate.

*4.1.1 Time Complexity.* For the tensor completion process, the time complexity to satisfy a user's query (i.e., obtain a rating prediction for a given product) is proportional to the $d$ entries in scaling vector $S_k$. Assuming fixed $d = 2$, the time complexity to evaluate TCA($A$) is given by the following theorem:

THEOREM 4.4 (TIME COMPLEXITY). *The time complexity for the RS procedure on tensor A is $O(|\sigma(A)|)$. With $d = 2$, the time complexity of the RS procedure on matrix A is $O(|\sigma(A)|)$.*

In the following section, we conclude our analysis of the framework with consideration of how unit consistency implies a precise notion of *fairness* with respect to the relative influence of each user's ratings on the recommendations produced by a UC recommender system.

## 4.2 Fairness and Robustness

Applying a notion of "fairness" to an algorithmic process is inherently challenging for a variety of reasons. The primary one is the introduction of an implicit moral judgment, i.e., defining what is "fair" versus "unfair," which may not be universally accepted. The second is constructing a definition that is unambiguous and computationally tractable to both satisfy and check. In the case of recommender systems, there are obvious concerns about how user attributes such as gender and ethnicity may potentially be misused. This is further complicated by the fact that some products are specifically designed to address a need of a particular group that may have previously been unmet due to the minority status of that group [5]. Some concerns may be addressed trivially by removing certain attributes from use by the RS formula, but beyond that, the algorithm will most likely require human-imposed constraints that may or may not be efficiently satisfied.

For our purposes, we focus on fairness in terms of ensuring that the relative preferences of all users are weighted equally by the system. It is impossible, of course, to prevent users from entering false information that does not reflect their true preferences (*garbage-in-garbage-out*). However, a given RS algorithm may be susceptible to bias that causes the good-faith ratings of one user to be given more weight/influence than the good-faith ratings of another user. For example, suppose Alice rates all 8,000 films in a particular database (which is presumed to roughly span all genres) and gives an average rating of 86/100, whereas Bob rates the same 8,000 films with an average rating of 43/100. Now further assume – purely to emphasize a point – that Alice's rating for each film is exactly double that given by Bob. Is it reasonable to hypothesize that Alice and Bob have nearly identical tastes in film but apply different absolute units of measure when entering their subjective ratings? Said another way, if Alice were to rate a new film as 80/100, is it reasonable to expect that Bob will likely give it a rating of approximately 40/100? The answers to these questions are implicitly "yes" according to the unit-consistency criterion. More intuitively, it may be hypothesized that Bob is more conservative in his ratings than Alice, e.g., because he may be reluctant to give the maximum rating of 100/100 to any film because he foresees the possibility of a future film he might like even more, while Alice is less concerned about that prospect.

There are many equally-defensible ways to treat the given example of Alice and Bob for purposes of developing a recommender system. However, not all of the resulting algorithms can ensure that the relative weights given to their respective sets of preferences are equal when computing recommendations for other users. For example, a system that minimizes squared prediction error will apply more weight to larger-magnitude ratings. Thus, in the case of the Alice-Bob example, Alice's ratings will tend to have more influence on recommendations to other users than those of Bob simply because her ratings tend to be larger in magnitude with respect to the mean.[6] Consequently, a squared-error minimizing system is susceptible to generating recommendations that are biased toward the relative preferences of some users over those of other users, e.g., potentially diminishing the contribution of a particular subset of users who may be more reserved in their assignment of ratings due to cultural influences.

To summarize, the UC criterion leads to recommendations that are invariant with respect to a global scaling of any row or column of the RS matrix, so a UC-based recommender system is less susceptible to biases of the kind described and thus provides a provable, albeit weak, form of fairness. More importantly, UC eliminates a means for users to intentionally scale their ratings so as to cause the system to give recommendations that are more aligned with their preferences. Returning to the Alice and Bob example, for a system that is *not* unit-consistent, e.g., minimizes squared error, Bob could increase his influence on the system to equal that of Alice by scaling his ratings by a factor of 2 – *or he could scale by a larger factor to have more influence than her*. A unit-consistent RS is robust with respect to manipulation in this way because the relative ratings of each user are given equal weight by the RS system in the generation of predicted ratings. The following theorem formalizes this property:

THEOREM 4.5. *Let $A$ be a tensor, RS be the recommender system obtained from TCA(A), and $RS(u,p)$ denote the recommendation result for user $u$ and product $p$ in RS. For a given user $u$ and a sub-tensor $A_i$ representing their subset of ratings for different products, let $T_{d-1} = \{T_{\{u\}}\}_{u \in [d]}$ be a scaling vector applied by user $u$ to the RS such that $T_{\{u\}} > 0$ and $T_{\{u'\}} = 1$ for $u' \neq u$, resulting in a new recommender system $RS_{new}$. Then for any user $u' \neq u$ and product $p$, we have $RS_{new}(u',p) = RS(u',p)$.*

We conclude this section with an example that is intended to provide stronger intuition about both the fairness property and the definition of full-support required for solution uniqueness within the UC framework. Consider a large database of users and films in which Alice gives a rating of 10/10 for every film she rates. Based on the unit-consistency property, it can be inferred that recommendations to other users should be the same if she had given a rating of 1/10 for every film. This is true because she has no other ratings from which to establish an absolute measure of relative preference due to lack of full support. Thus, her all-equal ratings will have no effect on system recommendations. This could be regarded as a bias against users who choose only to submit ratings for films they thoroughly like (or, alternatively, strongly dislike), but at least it is a property of the system that can be identified and rigorously analyzed directly from an intuitive understanding of the unit consistency property.

## 5 CONSIDERATIONS ON RS PERFORMANCE ASSESSMENT

In this section we discuss various issues relating to the evaluation of RS performance. We then provide empirical corroboration for theoretically-proven UC properties relating to fairness and robustness.

---

[6]It may be tempting to suggest that the RS translate ratings to have zero mean before processing, or possibly define the rating interval to be something like -10 to +10, but doing so would then give greater weight to users who preferentially only rate films they either strongly like or strongly dislike. It is possible that the ratings of such users may in fact tend to be good predictors for other users, but if greater RS weight is given to ratings at the extremes, then that becomes a potentially exploitable property of the system. As has been emphasized, any RS can be manipulated, but it is preferable to have properties that are relatively transparent, as opposed to being subtle mathematical artifacts resulting from optimizing an arbitrary measure of error.

## 5.1 Descriptive vs. Prescriptive Models

A question that can arise in the development of a model for a system is whether it is intended to be *descriptive* or *prescriptive*. A descriptive model attempts to statistically describe the state or behavior of a system of interest. If the system is stationary in the statistical sense that its fundamental properties do not change dynamically over time, then a high-fidelity descriptive model of the system has the potential to permit high-fidelity predictions to be made about unobserved attributes of the system, e.g., how a particular user is likely to rate a particular product. As will be discussed later, however, descriptive modeling of nonstationary attributes of a system can lead to unexpected issues.

A prescriptive model is somewhat analogous to the laws of physics in the sense that it is defined in terms of fundamental properties that are inferred to hold universally. Making the leap from an observed statistical property to a universal "law" allows strong inferences to be made in low-information contexts by imposing constraints on the possible state of the system being observed. For example, if measurements are taken of an assumed linear system, then observed deviations from linearity must be interpreted as noise. A descriptive model, by contrast, has no *a-priori* means for distinguishing signal from noise, so it is strongly susceptible to constructing an over-parameterized representation until sufficient information is processed over time to gradually smooth out any spurious nonlinear artifacts so that a simpler representation becomes possible.

Some intuition can be garnered from consideration of a simple descriptive model that is intended to determine the degree and coefficient values of a polynomial that accurately describes a scalar system of points based on noisy observations. Figure 1 provides an example involving 10 points, subjected to Gaussian noise, observed from a model that is defined by a cubic polynomial. In other words, a cubic polynomial defines the ground truth behavior of the underlying model. As can be seen in Figure 1, higher-degree polynomials can always provide better fits to the observed noisy points according to any chosen measure of error than lower-degree ones, *even if one of the lower-degree polynomials corresponds to the actual ground truth system*. This is because the extra hyperparameters (degrees of freedom) are able to provide fits to purely random noise artifacts that have nothing to do with the actual system of interest.

More generally, a prescriptive model can be viewed as the high-information limit of an idealized descriptive model. We have constructed a prescriptive model by taking unit consistency to be a fundamental property of recommender systems. However, the larger message of this paper is that a prescriptive model should be preferred in order to establish rigorous performance properties, e.g., to provably guarantee that certain undesirable behaviors cannot occur. This is not typically possible for descriptive models because, as already mentioned, they are inherently susceptible to significant transient biases resulting from low-information statistical artifacts. Of potentially greater concern in the context of recommender systems is vulnerability to feedback amplification.

*5.1.1 Feedback Amplification.* It is important to recognize that a recommender system is intended to function in a feedback loop in which its recommendations are intended to influence the users of that system. Specifically, it is expected/hoped that recommendations to users will lead to their purchasing of products they will tend to rate highly. Consequently, users will tend to experience more products they like and fewer they dislike. Over time, therefore, it should be expected that the ratio of high to low ratings will increase, i.e., the mean rating becomes a nonstationary statistic.

Because a recommender system is intended to influence user ratings, it is difficult to imagine any statistical attribute of collected data that will not be affected, and this creates a potentially complex feedback loop. This should not be a problem for a prescriptive model if it is based on constraints and properties that are truly fundamental within the problem domain. However, the situation is very different for a descriptive model that is based on lagging statistical estimates of a dynamical system that it is directly influencing. While the provable properties of a prescriptive model will continue to hold, it may be difficult to predict the effects of feedback amplification on the performance of a descriptive model, e.g., an AI-based system whose performance may be entirely opaque beyond its optimization of a given performance metric.

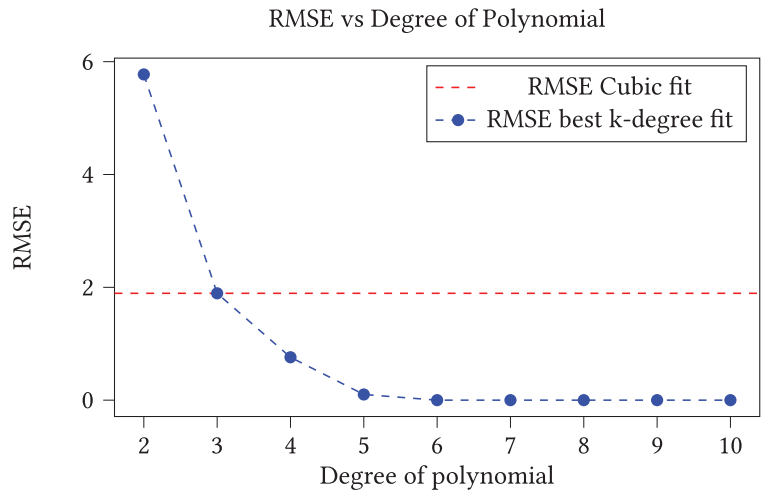

Fig. 1. Noisy observations of a simple system were simulated as set of 10 scalar values sampled from a cubic polynomial and then perturbed by zero-mean Gaussian noise. In other words, the ground truth model for the observations is a cubic polynomial. The red dashed line shows the RMSE for the best cubic polynomial fit to the points, and thus represents the best possible parameterized model that can be constructed from the available observations. The horizontal axis shows reduction in RMSE obtained by increasing the polynomial degree of an assumed model. With sufficiently high degree, it is always possible to perfectly fit a given set of observations, i.e., with zero error, but the apparent improvement must be spurious because the cubic model represents the best possible explanation.

As an example, consider a trivial discretionary attribute of a class of product that has virtually no actual significance to users but is initially deemed by a descriptive model to have a small – *though entirely spurious* – positive correlation with user preference. If there are many variants of the product that the model determines will be rated highly by users, then a slight bias toward recommendations for variants with the extraneous attribute will tend to lead to a disproportionate number of high ratings for those with that attribute. In other words, the number of products sold with the extraneous attribute will be biased toward those with other attributes that actually are preferred by users. This of course will lead to an increasingly strong correlation between the extraneous attribute and high ratings by users.[7] The consequences in this example may not seem particularly concerning, but consider similar biases involving variables with social implications, e.g., relating to race or gender.

One intuitive approach for avoiding feedback effects is to construct a descriptive model that is not subsequently updated based on information that it has influenced. In other words, a model with entirely static hyperparameters could be created based on a large initial training corpus, and no dynamic updating is applied to it after going live. The problem, of course, is that small biases arising from the limited initial training set will persist indefinitely because there will be no opportunity for them to be "filtered out" through the processing of additional information over time. This means that recommendations will perpetually maintain those biases and, through repeated exposures, may tend to steer user sentiments toward those biases.

---

[7]Analysis of this biased trend might even motivate producers to preferentially include the extraneous attribute in new variants of the product. In fact, user-experienced associations between the attribute and the high-quality variants of the product recommended by the system may actually lead users to develop a preference for the attribute!

## 5.2 Fairness and Robustness

Notions of "robustness" and "fairness" are challenging to define in a manner that allows for rigorous guarantees to be made. For example, a user can certainly affect a system by entering spurious ratings, and they may be able to do so in a way that is intended to have a specific effect on a particular product, e.g., improve or reduce its average rating, but no system can be immune to that. More specifically, no RS system can distinguish whether a given rating actually reflects the true opinion of the user. Under some conditions it may be possible to discern statistical evidence of rating manipulations that are intended to exploit some property of a given RS, and suspect users and ratings then may be removed as a remedy, but there can be no general mechanism for detecting false ratings. With this in mind, we mention two candidate definitions of fairness that can be enforced:

> **Scale-Fairness Property**: *A recommender system is said to be fair if its recommendations are invariant to a scaling factor applied to all of a user's ratings.*
>
> **Shift-Fairness Property**: *A recommender system is said to be fair if its recommendations are invariant to the addition of a constant value to all of a user's ratings.*

By definition, the UC method guarantees the scale-fairness property.[8] Here we provide a simple empirical test demonstrating that the predicted ratings from conventional rank-restricted SVD methods – *and, further, their rank orderings* – are sensitive to even a single user scaling her/his ratings. In this experiment, the ratings of a single user in the MovieLens1M dataset (specifically, user 4360) are scaled by 5/4. As can be seen in Figure 2, the UC method is unaffected. However, for an SVD algorithm [33] (labeled "SVD"), and a custom variant that is tailored for ranking [8] (labeled "Ranking SVD"), the top-N ordered sets of recommendations to users are significantly affected. For example, the top 5 recommended films given to over 250 users changed as a result of a single user scaling her/his set of ratings.

Consensus Ordering and Scale-Fairness are rigorous properties guaranteed by UC, and the results of Figure 2 provide empirical corroboration of those properties. Figure 3, by contrast, directly assesses the admissibility of all three methods according to our consensus order criterion. Specifically, the test involves the inclusion of three films: X, Y, and Z, all of which were given ratings 3, 2, and 1, respectively, by half the users, while none of the remaining users had rated any of them. According to the proposed admissibility criterion, any *reasonable* RS should respect that relative rank ordering in its recommendations to users who have not rated any of the three films, i.e., it would not make sense for any RS to recommend Y over X to a new user when all users who have seen the films have unanimously preferred X over Y and Y over Z. As can be observed from Figure 3, the UC method satisfies the criterion, whereas the two SVD methods commit significant violations until more than 90% of the singular values are retained. Different RS robustness and/or fairness properties may exist for the SVD, but Figure 3 suggests that it may be challenging to identify and prove such properties when only a subset of the singular values are retained (though it may be possible to determine a bound on the magnitude of perturbations based on a function of the non-retained singular values).

---

[8]We note that the logspace normalization step of our algorithm provides a mechanism for obtaining zero-sum rows and columns without imputing values to unfilled entries, and this is what motivates the Shift-Fairness definition. Shift-consistency will be the focus of future work [25] as a means to avoid potential multiplicative sensitivity to highly discretized ratings, e.g., the MovieLens 1 to 5 scale.

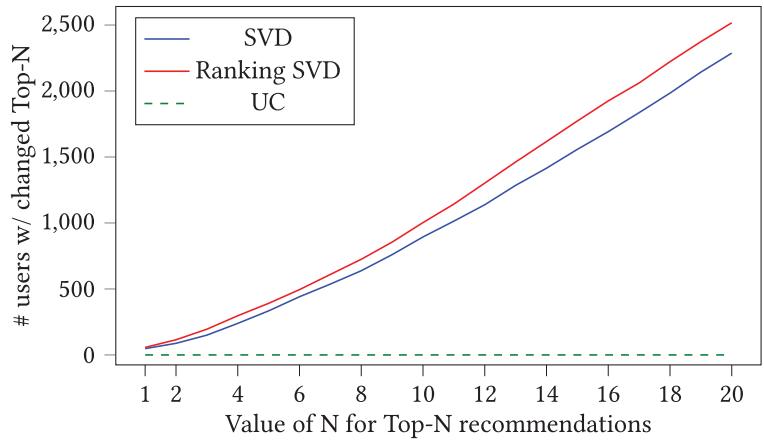

Fig. 2. This figure shows the effect of a single user labeled 4360 in the MovieLens1M dataset scaling their ratings by a factor of 5/4 so that their highest rating of 4 becomes the maximum rating value of 5. More specifically, we consider the extent to which such a scaling affects the rank ordering of recommendations made to other users. The horizontal axis represents the number of changes in the top-N, and the vertical axis represents the number of users whose top-N ranked recommendations from the system changes as a consequent of the single user scaling their ratings. For example, between 500 and 1,000 users would see one or more changes to their top-10 recommendations from the system.

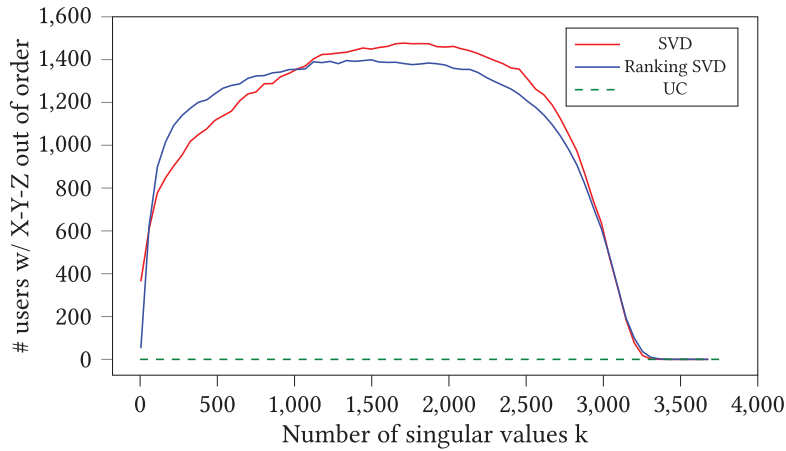

Fig. 3. This test used a subset of 3,604 films from the MovieLens dataset, to which films X, Y, and Z were added. Half of the users rated all three films as 3, 2, and 1, respectively. The remaining control set of users had no ratings for any of the three films. The results show that all UC recommendations given to the control set respected the unanimous rank-order of preference of users who had rated all three films, whereas neither of the SVD-based methods did so.

### 5.3 Equality Bias

RS algorithms that maintain discretized information, e.g., estimated rating values that are rounded to integers, will be forced to determine – either implicitly or explicitly – the order in which recommendations of equal "quality" are presented to the user. If this order is based on some form of lexicographic order, or the time at which each product was added to the system, or any other arbitrary deterministic ordering criterion, then some products will strongly benefit from consistently being ranked higher than other products of equally-assessed quality when presented to users. Simple randomization is therefore warranted to mitigate such biases. Randomization in the non-equal case may also be warranted for systems (e.g., $\epsilon$-admissible ones) when the rank-ordered list of recommendations includes numerically similar consecutive items that cannot be confidently regarded as truly rank-distinguishable.

Our framework is susceptible to this issue depending on how predicted ratings are transformed to the set of allowable rating values, e.g., integers in a fixed range, which should be done in practice. However, if the real rating is associated with its transformed rating, then equality of transformed ratings can be resolved unambiguously by ranking them according to the UC raw predicted ratings.

### 5.4 Complexity Considerations

The SVD can be thought of as a transformation of a matrix to a nonnegative diagonal canonical form. It is arguably the most versatile and general linear algebra tool, but it may be regarded as overkill for RS applications because its rotation-consistent properties do not naturally apply to the RS problem. In terms of computational complexity for an $m \times n$ matrix with $k$ singular values, its construction takes $O(mnk)$ time, or can be approximated in $O(mn \log(k))$ time [9]. This corresponds to the preprocessing complexity for an SVD-based RS system in the case of a dense matrix, i.e., having little or no sparsity. The UC algorithm, by contrast, takes strictly $O(mn)$ time with low computational overhead in the dense case, and it is linear in the number of filled entries in the sparse case, which is trivially optimal.[9]

### 6 DISCUSSION AND FUTURE WORK

A key contribution of this paper is the proposal of the consensus-order admissibility condition for proposed recommender systems, which says that if all users agree on a rank-ordering of a set of products, then recommendations (entry completions) must respect that ordering. Our position is that if a given system is not guaranteed to satisfy this basic intuitive expectation, then whatever it does is not aligned with what any reasonable person should expect of a recommender system. For example, if all users are unanimous in their rank-ordering of a set of products, and that includes unanimous agreement that product *A* is preferable to product *B*, then what should be concluded about a given system that recommends *B* over *A* to a new user? We suggest that such a violation of consensus ordering naturally raises questions of bias that directly relate to notions of fairness, inclusivity, and *trust*. How can users trust a system that recommends *B* over *A* when all users prefer *A* over *B*? Is it due to unintended bias or nefarious manipulation? This may be of direct concern to users from historically marginalized groups, especially if the products in question are of particular relevance to them.

Our approach to the assessment of RS candidates is a significant departure from the current state of affairs in which systems are typically only compared on a small number of specific datasets using generic performance measures, e.g., RMSE or MAE, and thus provide little or no basis for drawing general inferences about how they might perform on different datasets. In fact, the literature indicates that most systems perform comparably on standard benchmarks, with different systems performing better than others depending on the benchmark. This tends to suggest that most systems effectively exploit most of the available first-order information. Given the relatively small differences in performance, one can imagine the possibility that hyperparameter tuning might make a difference in a system's relative performance ranking on a fixed set of datasets, thus further reducing the interpretative significance of, e.g., simple best-to-worst rankings according to RMSE [24].

---

[9]The case of an $n \times n$ matrix of full rank may provide additional intuition. In this case the SVD has $O(n^3)$ complexity while UC has optimal $O(n^2)$ complexity, i.e., is linear in the number of elements in the $n \times n$ matrix.

The most significant contribution of this paper is a fully-rigorous RS framework that demonstrates how the admissibility criterion can be provably satisfied. Specifically, we have shown that the imposition of a unit-consistency constraint on the generation of recommendations is sufficient to guarantee satisfaction of the consensus-ordering property. We propose that our general approach of imposing a constraint (projecting to a restricted subspace) such that a required property is guaranteed (e.g., consensus-ordering), can be applied to arbitrary blackbox systems,[10] e.g., AI-based, that otherwise would not be amenable to any kind rigorous performance analysis beyond that of generic metrics such as RMSE under specific testing conditions. To emphasize more strongly, we suggest that no form of AI alignment can exist unless operations are restricted to a rigorously defined space in which desired properties are provably guaranteed. Our general position is that a system either has provable[11] properties sufficient to establish trust, or it is not trustworthy.

Generic measures of error such as RMSE and MAE can prove useful for selecting among a set of *admissible* RS candidates. In fact, if one candidate system performs better across a variety of generic metrics when tested on application-relevant datasets, then it may be hypothesized that some aspect of the candidate's structure captures some unknown aspect of that particular RS problem domain. Of course, this presumes that candidates are not specifically tailored to optimize those generic measures. What is most important to recognize, however, is that good RS performance according to a generic domain-independent measure of performance (such as RMSE) is not necessarily sufficient for the RS to be *trusted* to provide properties such as fairness and robustness. This is especially true when results for one-shot testing are implicitly assumed to reflect RS behavior when applied in a live feedback loop with users. We believe that the imposition of projective-space constraints will be needed to ensure that complex descriptive RS systems, e.g., AI-based, possess desired properties.

In summary, our philosophical approach to the recommender system problem has been to identify properties that should be expected of any *reasonable* solution, and consensus ordering is such a property. We have shown that unit consistency is sufficient on its own to yield a unique solution that satisfies the consensus ordering property. This is surprising because the recommender system problem seems at first glance to demand unknowable information deriving from individual human psychology, thus appearing to be unavoidably within the domain of AI. Our examination of the recommender system problem shows that recognition of a few constraints on the solution space can sometimes replace profound mystery with simple clarity.

## APPENDICES

### A GENERAL CASE DEFINITION

Here we state general definition for any $k$-dimensional subtensor and scaling vectors.

- $A_i$ is the $i^{th}$ element of the ordered set of all $k$-dimensional subtensors of $A$, where each subtensor is equal to $A$ but with a distinct subset of the $d - k$ extents restricted to 1. If vector $\vec{\alpha}$ satisfies $A(\vec{\alpha}) \in A_i$, we write $\vec{\alpha} \in A_i$.
- $S_k$ is a strictly positive-valued vector of length equal to the number of $k$-dimensional subtensors of $A$, with $S_{k,i}$ denoting the $i^{th}$ element of $S_k$.

---

[10]Upcoming works examine this approach with blackbox systems in more detail [25, 26].
[11]Empirical results may provide a useful average-case characterization of a system's behavior, but they are not sufficient for allaying *worst-case* concerns.

By Definition 3.2, each coefficient $S_{k,i}$ scales the corresponding subtensor $A_i \in \mathcal{A}$ to derive tensor $A''$, giving us the equivalent formula.

$$A''(\vec{\alpha}) \equiv A(\vec{\alpha}) \cdot \prod_{i:\vec{\alpha} \in A_i} S_{k,i} . \tag{19}$$

This formula gives us the general formulation of the scaling tensor problem for any arbitrary dimensional number $k$.

## B MAIN ALGORITHMS FOR GENERAL CASE

We present two algorithms for obtaining a unique scaled tensor $A' = CSA(A, k)$ and the tensor completion algorithm $TCA(A, k)$ for any value $k$.

---
**ALGORITHM 3:** General TCA

**Input**: $d$-dimensional tensor A and $k$.
**Output**: $A'$
**Function** TCA$(A, k)$:
- **Step 1: CSA process.**
  $S_k \leftarrow$ CSA$(A, k)$
- **Step 2: Tensor completion process**:
  **for** $\vec{\alpha} \in \overline{\sigma}(A)$ **do**
  $\quad A'(\vec{\alpha}) \leftarrow \prod_{i:\vec{\alpha} \in A_i} S_{k,i}^{-1}$ .
  **return** $A'$.

---

**ALGORITHM 4:** General CSA for tensor

**Input**: $d$-dimensional tensor $A$.
**Output**: $A''$ and scaling vector $S_k$.
**Function** CSA$(A, k)$:
- **Step 1: Iterative step over constraints**: Initialize $count \leftarrow 0$, variance variable $v \leftarrow 0$, and let $p$ be a zero vector of conformant length. Let $a$ be the logarithm conversion of $A$, i.e., all known entries are replaced with their logs.
  **for** *each subtensor $A_i$ with index i* **do**

$$\rho = -[|\sigma(A_i)|]^{-1} \sum_{\vec{\alpha} \in \sigma(A_i)} a(\vec{\alpha})$$
$$a(\vec{\alpha}) \leftarrow a(\vec{\alpha}) + \rho, \; v \leftarrow v + \rho^2, \quad \text{for} \quad \vec{\alpha} \in \sigma(A_i) \tag{20}$$
$$s_{k,i} \leftarrow s_{k,i} - \rho$$

- **Step 2: Convergence**: If $v$ is less than a selected threshold $\epsilon$, then exit loop. Otherwise, set $count \leftarrow count + 1$ and return to step 1.
  **return** $A'' = exp(a)$ and $S_k = exp(s_k)$.

---

The main algorithms $TCA(A)$ and $CSA(A)$ assume $k = d$-1 for $TCA(A, k)$ and $CSA(A, k)$. For the subsequent sections of the Appendix, we focus on the general theorems relating to these two algorithms.

## C GENERAL CANONICAL SCALING PROBLEM

Here we formulate the general form of the canonical scaling problem in the real space and in the log space:

**Canonical Scaling Problem (CSP):** Find the $d$-dimensional tensor $A' \in \mathbb{R}_{>0}^{n_1 \times \cdots \times n_d}$ and a positive vector $S_k$ such that $A'' = A *_k S_k$ and the product of the known entries of each $i^{th}$-index $k$-dimensional subtensor $A_i''$ is 1.

**Log Canonical Scaling Problem (LCSP):** The problem can be formulated as follows: Find the $d$-dimensional tensor $a''$ and a vector $s_k$ such that $a''(\vec{\alpha}) \equiv a(\vec{\alpha}) + \sum_{i:\vec{\alpha} \in A_i} s_{k,i}$, $\forall \vec{\alpha} \in \sigma(A)$, and each $i^{th}$-index $k$-dimensional subtensor $a_i$ with the sum of its known entries equal to zero.

$$\sum_{\vec{\alpha} \in \sigma(A_i)} a''(\vec{\alpha}) = 0 \quad \forall A_i \in \mathcal{A} \tag{21}$$

### C.1 Existence of Solution

We established a general existence proof of the **LCSP** in [27] based on [20] and [32]. Here we provide a sketch of the base case. Let $a \in R^p$, and matrix $C \in R^{p \times q}$ be the original convex optimization problem of finding a vector $a' \in R^p$ and $\omega \in R^q$ such that

$$a'^T = a^T + \omega^T C \tag{22}$$

and

$$Ca' = 0 \tag{23}$$

It is proven in [32] that this program is equivalent to the following optimization problem, for which the properties of uniqueness and existence can be established:

$$\min 2^{-1} \sum_{i=1}^{p} (x_i - a_i)^2 \tag{24}$$
$$\text{subject to} \quad Cx = 0.$$

In [20] and [32], the equivalency of the aforementioned two optimization problems guarantee the existence of a unique solution. Based on this, we demonstrate that LCSP is equivalent to the following **Convex Optimization Problem (COP)**; hence, we can deduce the existence of a tensor $A'$ that satisfies CSP and has $\sigma(A) = \sigma(A')$ such that for the log tensor $a$ of $A$:

$$\underset{x \in \mathbb{R}^{n_1 \times \cdots \times n_d}}{\text{minimize}} \; 2^{-1} \sum_{\vec{\alpha} \in \sigma(A)} (x(\vec{\alpha}) - a(\vec{\alpha}))^2 \quad \text{subject to} \sum_{\vec{\alpha} \in \sigma(A_i)} x(\vec{\alpha}) = 0, \quad \forall \text{ subtensor } A_i. \tag{25}$$

It should be noted that the set of all possible elements $s_{k,i}$ for equation $Cx = 0$ maps onto $\omega$ as $\omega_i = s_{k,i}$. With our slight abuse of notation, we perform a tensor-flattening matching from tensor $a$ to vector $a$ via the mapping $a(\vec{\alpha}) \mapsto a(J(\vec{\alpha}))$ for

$$J(\vec{\alpha}) = 1 + \sum_{s=1}^{d} \left( (\alpha_s - 1) \prod_{m=1}^{s-1} n_m \right)$$
$$= \alpha_1 + (\alpha_2 - 1)n_1 + \cdots + (\alpha_d - 1)n_1 \cdots n_{d-1}$$

Note that vector $a$ agrees with tensor $A$ in terms of nonzero elements. In the COP problem, we can match the flattening vector $x$ with vector $a$ via unfolding of the aforementioned mapping. Matrix $C$ becomes the medium to suit $\omega^T$ and $x$ with the conditions in COP and LCSP. The form of $C$ is established such that it can select the corresponding scaling elements for $\omega^T C$ at position $J(\vec{\alpha})$ and select nonzero elements of a subtensor by the index $i$ in $Cx$. The matrix $C$ is defined as

$$C_{i,J(\vec{\alpha})} \equiv \begin{cases} 1 & \text{if } \vec{\alpha} \in \sigma(A_i) \\ 0 & \text{otherwise.} \end{cases} \tag{26}$$

The ordering of columns in $C$ follows the ordering of both $a$ and $a'$. This ordering implies that for each $\vec{\alpha}$ in $\sigma(A)$, the column $J(\vec{\alpha})$ considers only dimensional positions in $\vec{\alpha}$.

$$(Cx)_i = \sum_{J(\vec{\alpha})} C_{i,J(\vec{\alpha})} x(J(\vec{\alpha})) = \sum_{\vec{\alpha} \in \sigma(A_i)} a(\vec{\alpha}). \tag{27}$$

Thus, $Cx = 0$ is equivalent to

$$\sum_{\vec{\alpha} \in \sigma(A_i)} a(\vec{\alpha}) = 0. \tag{28}$$

Likewise, equation $\omega^T C$ satisfies

$$[\omega^T C]_{J(\vec{\alpha})} = \sum_{i: \vec{\alpha} \in A_i} \omega_i C_{i,J(\vec{\alpha})} = \sum_{i: \vec{\alpha} \in A_i} s_{k,i} \tag{29}$$

Thus, the convergence theorem for the LCSP follows from the aforementioned optimization problem (24). If we allow a matrix transformation of LCSP, the problem can be formulated as the following system of linear equations:

$$\begin{cases} C^T \mathbf{s} + I_{\sigma(A)} \mathbf{a}' = \mathbf{a} \\ C \mathbf{a}' = 0 \end{cases} \Leftrightarrow \begin{bmatrix} C^T & I_{\sigma(A)} \\ 0 & C \end{bmatrix} \begin{bmatrix} \mathbf{s} \\ \mathbf{a}' \end{bmatrix} = \begin{bmatrix} \mathbf{a} \\ 0 \end{bmatrix}$$

This equation can be represented as an upper triangular block matrix equation and can be solved in back-substitution by Gaussian elimination in $O(|\sigma(A)|^3)$ time and in back/forth substitution in $O(|\sigma(A)|^2)$ time. These methods can be applied to obtain the desired scaling transformations, but they incur much higher computational complexity because they do not exploit the simpler structure of the problem, e.g., the finding of a diagonal scaling versus an arbitrary linear transformation of a matrix. Our algorithm utilizes diagonal scaling to achieve a time complexity of $O(|\sigma(A)|)$ per iteration for an overall time complexity of $O(|\sigma(A)|)$ for fixed $d$.

### C.2 Uniqueness of General LCSP

We know that if there exists a solution $x$ such that $Ax = 0$, and vector $\omega$ such that $A\omega = 0$, then $x + \omega$ is another solution. This leads to the following theorem regarding solution uniqueness:

THEOREM C.1 (UNIQUENESS OF A''). *There exists at most one tensor $A'$ for which there exists a strictly positive vector $S_k$ such that the solution $(A'', S_k)$ satisfies CSP. Furthermore, if a positive vector $T_k$ satisfies*

$$\prod_{i: \vec{\alpha} \in A_i} T_{k,i} = 1 \quad \forall \vec{\alpha} \in \sigma(A), \tag{30}$$

*then $(A'', S_k \otimes T_k)$ is also a solution, where $\otimes$ is the element-wise product.*

PROOF. The first part of the theorem is formally proven in [27]. For the second part, we consider the following condition.

$$\sum_{i: \vec{\alpha} \in A_i} t_{k,i} = 0 \quad \forall \vec{\alpha} \in \sigma(A), \tag{31}$$

for which the LCSP condition becomes:

$$a''(\vec{\alpha}) \equiv a(\vec{\alpha}) + \sum_{i: \vec{\alpha} \in A_i} (s_{k,i} + t_{k,i}), \quad \forall \vec{\alpha} \in \sigma(A). \tag{32}$$

Thus, the vector form $s_k + t_k$ in logarithm space is equivalent to $S_k \otimes T_k$ in the real space. □

## D GENERAL UNIQUENESS PROOF

Here we restate the uniqueness theorem for general case, in which its corollary for the case $k = d-1$ is Theorem 3.6.

THEOREM D.1 (UNIQUENESS). *The result from $TCA(A, k)$ is uniquely determined even if there are distinct sets of scaling vectors $\{S_k\}$ that yield the same, unique, $A''$ from $CSA(A, k)$.*

To facilitate the subsequent theorem and proof, we introduce the following definition.

*Definition D.2.* For $\pi \subseteq [d]$ and $\vec{\alpha} \in \mathbb{Z}^d$, a vector $\vec{\alpha}_\pi \in \mathbb{Z}_{>0}^k$ for $k \leq d$ is considered a $k$-dimensional subvector of $\vec{\alpha}$ if $\vec{\alpha}_\pi = (\alpha_{\pi_1}, \ldots, \alpha_{\pi_k})$. We define the set of those $k$-dimensional subvectors as $V_k(\vec{\alpha})$.

Using this definition, we obtain the following proof regarding uniqueness of the recommendation/entry-completion result.

PROOF. For $A' = TCA(A, k)$, since entry $\vec{\alpha} \in \sigma(A)$ has $A'(\vec{\alpha}) = A(\vec{\alpha})$ by Theorem 3.4, we need only prove uniqueness of any completion $\vec{\alpha} \in \bar{\sigma}(A)$. From the uniqueness result of Theorem 3.4, $TCA(A, k)$ admits two distinct scaling vectors $S_k$ and $S'_k$ that yield the same, unique, $TCA(A, k)$. From Definition 3.5, there exists $2^d - 1$ vector $\vec{\alpha}' \in H(\vec{\alpha}, \vec{s})$ for some fixed vector $\vec{s}$. From Theorem 3.4, the scaling vector $S'_k$ equals $S_k \circledast T_k$ is equivalent to

$$\prod_{i:\vec{\alpha}' \in A_i} T_{k,i} = 1 \quad \forall \vec{\alpha}' \in H(\vec{\alpha}, \vec{s}). \tag{33}$$

We now show that

$$A'(\vec{\alpha}) = \prod_{i:\vec{\alpha} \in A_i} S_{k,i}^{-1} = \prod_{i:\vec{\alpha} \in A_i} S'^{-1}_{k,i} \tag{34}$$

or equivalently from Theorem 3.4

$$\prod_{i:\vec{\alpha} \in A_i} T_{k,i} = 1. \tag{35}$$

Without loss of generality, we consider the case $d \equiv 0 \pmod{2}$, and the other case can be proven similarly. We define two sets $G_0$ and $G_1$ by the following. Except for $\vec{\alpha}' = \vec{\alpha} + \vec{s}$, we divide $2^d - 2$ remaining vectors $\vec{\alpha}'$ into two groups. Then for $m \in \{0, 1\}$, $G_m$ is the set of $\vec{\alpha}' = \vec{\alpha} + \vec{\delta} \cdot \vec{s}$ such that the number of $\delta_i = 0$ equals $m$ modulo 2. We also denote $G_m \cap A_i = \{\vec{\alpha}' \in G_m \mid \vec{\alpha}' \in A_i\}$. Then

$$\prod_{\vec{\alpha}' \in G_m} \prod_{i:\vec{\alpha}' \in A_i} T_{k,i} = 1 \Leftrightarrow \prod_i T_{k,i}^{|G_m \cap A_i|} = 1. \tag{36}$$

By Definition D.2, we consider $\vec{v} \in \bigcup_{\vec{\alpha}' \in H(\vec{\alpha}, \vec{s})} V_k(\vec{\alpha}')$ and a subset $\pi' \subseteq [k]$ as in Definition 3.1 such that $0 \leq k' \leq k$. Then the complement subset $\pi'^C \in [k] - \pi'$ gives $\vec{v}'_{\pi'} = \vec{\alpha}_{\pi'}$ and $\vec{v}'_{\pi'^C} = (\vec{\alpha} + \vec{s})_{\pi'^C}$. Consider case 1 when $0 < k' < k$, we form a fixed $\vec{\alpha}'$ from $\vec{v}$ by a new subset $\pi$ such that $\pi \subseteq [d]$ and $\pi' \subseteq \pi$ by a difference of $l$ elements, making $\pi$ has $k' + l \leq d$ elements. Then the complement subset $\pi^C \in [d] - \pi$ gives $\vec{\alpha}'_\pi = \vec{\alpha}_\pi$ and $\vec{\alpha}'_{\pi^C} = (\vec{\alpha} + \vec{s})_{\pi^C}$.

For $m \in \{0, 1\}$, if the number of elements of $\pi$ as $k' + l \equiv m \pmod{2}$, then $\vec{v}$ forms $\binom{d-k}{l}$ numbers of $\vec{\alpha}'$ that belongs to $G_m$. For subtensor $A_i$ and with the sum in between $0 \leq l \leq d - k$, $|G_m \cap A_i|$ equals $\sum_{k'+l \equiv m \pmod{2}} \binom{d-k}{l} = 2^{d-k-1}$ for any $m \in \{0, 1\}$. Thus,

$$\frac{T_{k,i}^{|G_1 \cap A_i|}}{T_{k,i}^{|G_0 \cap A_i|}} = 1. \tag{37}$$

If $k' = k$, we encounter the vector $\vec{\alpha}$ when forming $\vec{\alpha}'$. If $k' = 0$, we encounter the vector $\vec{\alpha} + \vec{s}$ when forming $\vec{\alpha}'$. Since we omit $\vec{\alpha}$ and $\vec{\alpha} + \vec{s}$ from $G_0$, $|G_0 \cap A_i| = 2^{d-k-1} - 1$ and $|G_1 \cap A_i| = 2^{d-k-1}$ in either case of $k'$. Thus,

$$\frac{T_{k,i}^{|G_1 \cap A_i|}}{T_{k,i}^{|G_0 \cap A_i|}} = T_{k,i} \tag{38}$$

and therefore

$$\frac{\prod_i T_{k,i}^{|G_1 \cap A_i|}}{\prod_i T_{k,i}^{|G_0 \cap A_i|}} = 1 \Rightarrow \prod_{i:\vec{\alpha} \in A_i} T_{k,i} \prod_{i:\vec{\alpha}' \in A_i} T_{k,i} = \prod_{i:\vec{\alpha} \in A_i} T_{k,i} = 1. \tag{39}$$

This equality implies that $A'$ is unchanged, and is thus uniquely determined. □

## E GENERAL UNIT-CONSISTENCY PROOF

Here we state the general UC theorem for any $k$.

THEOREM E.1 (UNIT-CONSISTENCY). *Given a tensor $A$ and an arbitrary conformant positive scaling vector $T_k$, $T_k *_k \mathrm{TCA}(A, k) = \mathrm{TCA}(T_k *_k A, k)$, where $T_k$ scales all $k$-dimensional subtensors of $A$ (with operator $*_k$ as defined in Definition* 3.2).

PROOF. Let $S_k \leftarrow \mathrm{CSA}(A, k)$. It can be shown that $A' = \mathrm{CSA}(T_k *_k A, k) = \mathrm{CSA}(A, k)$ for all $A$. With a slight abuse of notation, we assume all unknown entries of $A'$ are assigned the value of 1, i.e., $A'(\vec{\alpha}) = 1$ for $\vec{\alpha} \in \overline{\sigma}(A)$. The complete TCA process can then be defined as $\mathrm{TCA}(A, k) = S_k^{(-1)} *_k A'$, where $S_k^{(-1)} = \{S_{k,i}^{-1}\}$ is the inverse vector of $S_k$. Now, using the uniqueness Theorem 3.6, we can subsume the scaling vector $T_k$ into $S_k$ and deduce that

$$T_k *_k \mathrm{TCA}(A, k) = (S_k^{(-1)} \circledast T_k) *_k A' = \mathrm{TCA}(T_k *_k A, k). \tag{40}$$

□

15:26 T. Nguyen and J. Uhlmann[35] Zhao Song, David P. Woodruff, and Huan Zhang. 2016. Sublinear time orthogonal tensor decomposition. In *Proceedings of the 30th International Conference on Neural Information Processing Systems (NIPS'16)*. Curran Associates Inc., Red Hook, NY, USA, 793–801.

[36] Ruoyu Sun and Zhi-Quan Luo. 2015. Guaranteed matrix completion via nonconvex factorization. In *2015 IEEE 56th Annual Symposium on Foundations of Computer Science*. 270–289. https://doi.org/10.1109/FOCS.2015.25

[37] Anju Jose Tom and Sudhish N. George. 2019. Video completion and simultaneous moving object detection for extreme surveillance environments. *IEEE Signal Processing Letters* 26, 4 (2019), 577–581. https://doi.org/10.1109/LSP.2019.2900126

[38] Jeffrey Uhlmann. 2018. A generalized matrix inverse that is consistent with respect to diagonal transformations. *SIAM J. Matrix Anal. Appl.* 39, 2 (2018), 781–800. https://doi.org/10.1137/17M113890X arXiv:https://doi.org/10.1137/17M113890X

[39] Jeffrey Uhlmann. 2019. A Scale-Consistent Approach for Recommender Systems. (2019). https://doi.org/10.48550/ARXIV.1905.00055

[40] Wenhong Wang and Yan Lu. 2018. Analysis of the mean absolute error (MAE) and the root mean square error (RMSE) in assessing rounding model. *IOP Conference Series: Materials Science and Engineering* 324 (2018). https://doi.org/10.1088/1757-899X/324/1/012052

[41] Shengke Xue, Wenyuan Qiu, Fan Liu, and Xinyu Jin. 2018. Low-rank tensor completion by truncated nuclear norm regularization. In *2018 24th International Conference on Pattern Recognition (ICPR'18)*. 2600–2605. https://doi.org/10.1109/ICPR.2018.8546008

[42] Jingyu Yang, Yuyuan Zhu, Kun Li, Jiaoru Yang, and Chunping Hou. 2018. Tensor completion from structurally-missing entries by low-TT-rankness and fiber-wise sparsity. *IEEE Journal of Selected Topics in Signal Processing* 12, 6 (2018), 1420–1434. https://doi.org/10.1109/JSTSP.2018.2873990

[43] Guangxiang Zeng, Hengshu Zhu, Qi Liu, Ping Luo, Enhong Chen, and Tong Zhang. 2015. Matrix factorization with scale-invariant parameters. In *Proceedings of the 24th International Conference on Artificial Intelligence (IJCAI'15)*. AAAI Press, 4017–4024.

[44] Bo Zhang and Jeffrey Uhlmann. 2018. A Generalized Matrix Inverse with Applications to Robotic Systems. (2018). https://doi.org/10.48550/ARXIV.1806.01776

[45] Bo Zhang and Jeffrey Uhlmann. 2019. Applying a unit-consistent generalized matrix inverse for stable control of robotic systems. *Journal of Mechanisms and Robotics* 11, 3 (04 2019). https://doi.org/10.1115/1.4043371 arXiv:https://asmedigitalcollection.asme.org/mechanismsrobotics/article-pdf/11/3/034503/6759297/jmr_11_3_034503.pdf 034503.

[46] Bo Zhang and Jeffrey Uhlmann. 2020. Examining a mixed inverse approach for stable control of a rover. *International Journal of Control Systems and Robotics* 5 (2020). arXiv:https://www.iaras.org/iaras/filedownloads/ijcsr/2020/011-0001(2020).pdf.
Received 1 October 2022; accepted 21 May 2023ACM Transactions on Recommender Systems, Vol. 1, No. 3, Article 15. Publication date: August 2023.